\documentclass[aps,pra,twocolumn,showpacs,byrevtex]{revtex4}
\usepackage{graphicx}    
\usepackage{amsmath}
\usepackage{txfonts}

\newcommand{\unit}[1] {\,\mathrm{#1}}
\newcommand{\colNa} {\gamma_\mathrm{Na}}
\newcommand{\colbg} {\gamma_\mathrm{bg}}
\newcommand{\coltot} {\gamma_\mathrm{tot}}

\begin{document}
\title{Light-Induced Atomic Desorption for loading a Sodium Magneto-Optical Trap}
\author{Gustavo Telles}
\author{Tetsuya Ishikawa}
\author{Matthew Gibbs}
\author{Chandra Raman}
\email{chandra.raman@physics.gatech.edu}
\address{School of Physics, Georgia Institute of Technology, Atlanta, Georgia 30332, USA}
\date{\today}

\begin{abstract}
We report studies of photon-stimulated desorption (PSD), also known as light-induced atomic desorption (LIAD), of sodium atoms from a vacuum cell glass surface used for loading a magneto-optical trap (MOT). Fluorescence detection was used to record the trapped atom number and the desorption rate.  We observed a steep wavelength dependence of the desorption process above $2.6\unit{eV}$ photon energy, a result significant for estimations of sodium vapor density in the lunar atmosphere.  Our data fit well to a simple model for the loading of the MOT dependent only on the sodium desorption rate and residual gas density.  Up to $3.7 \times 10^7$ Na atoms were confined under ultra-high vacuum conditions, creating promising loading conditions for a vapor cell based atomic Bose-Einstein condensate of sodium.

\end{abstract}


\maketitle

\section{Introduction}

LIAD (light-induced atomic desorption) is a new tool for producing laser cooled samples of K, Rb and Cs \cite{anderson01,alexandrov02,atutov03,aubin05,klempt06}.  Its primary appeal is the non-thermal nature of the desorption process \cite{abramova84,meucci94}, which allows for fast switching--the gaseous vapor can be immediately quenched upon shut-off of the light.  Even in tiny quantities, residual vapor can dramatically reduce the efficiency of evaporative cooling to Bose-Einstein condensation (BEC), and LIAD has been shown to mitigate this problem \cite{shengwang04}.  Other solutions to the background vapor problem involve loading from differentially pumped secondary sources such as a dual MOT or slow atomic beam.  Typically these are more cumbersome approaches in comparison to a single vapor cell.

In this paper we report the use of LIAD to load a sodium MOT with up to $3.7 \times 10^7$ atoms in a pyrex vacuum cell.  Moreover, we demonstrate that such a MOT is compatible with vacuum limited lifetimes in the several second range.  We present measurements of the wavelength dependence of the LIAD effect, showing that it increases dramatically in the blue to ultraviolet region of the spectrum, consistent with a photoelectric threshold.

Photon-stimulated desorption (PSD) studies of neutral sodium are particularly important for a better understanding of the lunar surface and atmospheric composition. Sunlight irradiating the silicates present on the moon's surface creates a significant sodium and potassium vapor in the tenuous atmosphere, causing the moon to glow due to emission from the same vapor \cite{madey02}.  Optically induced desorption studies have been performed on small sodium clusters \cite{trager93}, Sapphire \cite{bonch90} and SiO$_2$ \cite{yakshinskiy99} surfaces, as well as thin organic films \cite{atutov03,meucci94,gozzini93,atutov99,marinelli06}.  Our measurements on atomic sodium are the first to use a MOT to probe PSD, and they may open up the possibility of using laser cooling techniques for chemical identification of lunar samples.

\begin{figure}[!t]
\centering
\includegraphics[width=0.5\textwidth]{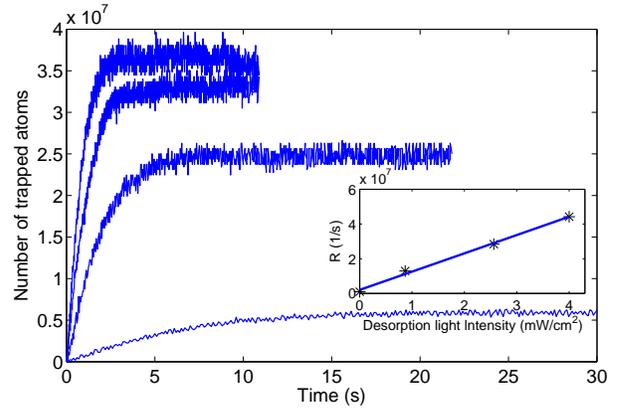}
\caption{LIAD assisted loading of a sodium MOT.  Atoms were captured from a transient vapor created by LIAD.  Shown are MOT loading curves for different light intensities using a blue high-flux LED ($455\unit{nm}$): $0, 0.9, 2.6$, and $4.0\unit{mW/cm^2}$. Simple exponential loading curves fit each data set and the inverse time constant $\gamma$ extracted from the fits is presented. We observe that the number of trapped atoms in the steady state increases monotonically with the LIAD intensity, saturating at high intensities. In the inset graph, the MOT loading rate is plotted as a function of the LIAD intensity, presenting a linear dependence, in very good agreement with a simple model described in the text.}
\label{fig:fitting}
\end{figure}

Apart from extraterrestrial applications, we believe our results will contribute to the widening use of vapor cell technology for producing better controlled Na MOTs in the range of tens of millions of atoms \cite{wippel01,Muhammad08}.  Indeed, a recent preprint reports a vapor cell based sodium BEC by rapid evaporation in a dipole trap, demonstrating that Zeeman slower technology is not required \cite{mimoun09} .  The sodium atom has many favorable properties for Bose-Einstein condensation, including a large ratio of elastic to inelastic scattering rates \cite{Moerdijk96}.  Its light mass contributes to large quantum mechanical effects, including the Bose-Einstein phase transition temperature, the separation of matter wave interference maxima, Josephson tunneling probability and atomic recoil shift \cite{pita03book}.  These phenomena are all of fundamental importance in addition to their significance for building practical atom optical devices.

\section{Experiment} \label{sec:experiment}

We used light from a single mode dye laser (Coherent 899, locked to near the $F=2\rightarrow F '=3$ hyperfine components of the D2 line of sodium at 589 nm by saturation spectroscopy) to make a standard retro-reflected three beam magneto-optical trap \cite{metc99}.  Prior to entry into the vacuum chamber the trap laser beam passes through a $1712\unit{MHz}$ electro-optic modulator (New Focus model 4461), which introduces about $20\%$ sidebands on the $F=1\rightarrow F'=2$ repumping transition.  Typically, $250\unit{mW}$ of total laser power was used from trapping, split into three independent beams of $20\unit{mm}$ diameter (laser power and beam diameters chosen to optimize the MOT), and traveling along the X,Y and Z directions.  A pair of quadrupole coils created the magnetic field gradient of 20 gauss/cm (at $I=15A$) along the vertical direction.

The vacuum chamber comprises a Pyrex cell of dimensions 45 $\times$ 45 $\times$ 150 $\unit{mm}$ attached to one face of a stainless steel 2 3/4" ConFlat cube (MDC).  The cube is also attached to a 20 l/s ion pump and titanium sublimation pump, and could be evacuated through an all-metal seal valve.  The titanium sublimation section was added to achieve UHV conditions (pressure below the lower limit of the vacuum gauge, $<1 \times 10^{-11}\unit{torr}$) and the experiments with HV (pressure $\approx 6\times 10^{-9}\unit{torr}$) were done without this additional pump.  The overall dimensions of the chamber were 50 $\times$ 55 $\times$ 52 $\unit{cm}$, rendering it light and easily removed for modifications or maintenance.

The sodium source was a set of alkali dispensers from SAES (SAES Alkamax metal dispensers, model NA/NF/1.5/12 FT 10+10, active length $12\unit{mm}$, threshold current $6\unit{A}$). Copper wires extending from a vacuum feedthrough were used to run current to the dispensers as well as to secure them about $1\unit{cm}$ inside the cell. The actual distance between the MOT and the dispenser could be varied from $15\unit{mm}$ to $70\unit{mm}$ by simply translating the vacuum chamber relative to the MOT center.  No significant variation of the number of trapped atoms was observed, and therefore during the experiments this distance was fixed around $40\unit{mm}$. For the HV experiments, one dispenser was fired at $5$ to $7\unit{A}$ during experimental runs to fill the cell with sodium. The MOT coil current was used to turn the trap on and off.  For UHV, we first fired the same dispenser at no more than $3\unit{A}$ for 30 minutes to coat the cell walls and waited for a few hours until the pressure recovered. In this case, a loading sequence was initiated by turning on the LIAD light a few seconds before the coil current was on in order to allow the LED to thermally equilibrate. In both cases, fluorescence from the MOT was recorded by a calibrated photodetector whose output current is proportional to the atom number. The photocurrent was recorded on an oscilloscope and comprised all of the data recorded and presented here.  The calibration between photodetector current $I_{ph}$ and atom number $N$ can be written as \cite{metc99}

\begin{equation}
N = \frac{1}{\Gamma_\mathrm{sc}} \frac{I_\mathrm{ph}/\mathfrak R_\lambda}{hc/\lambda} \frac{4\pi}{\Omega}
\end{equation}

where $\Gamma_\mathrm{sc} (I) = \frac{\Gamma}{2} s/(1+s+4 \delta^2/\Gamma^2)$ is the laser-intensity dependent scattering rate.  $s = I/I_{sat}$ is the saturation parameter with $I_{sat} = 6.3\unit{mW/cm^2}$ for Na.  $\delta = -15\unit{MHz}$ was the detuning and $\Gamma = 9.8\unit{MHz}$ is the natural linewidth.  The other parameters are the responsivity of the Si photodector $\mathfrak R_\lambda = 0.38 \unit{A/W}$ at $\lambda = 589 \unit{nm}$, and $\Omega = 0.1\unit{sr}$ the solid angle covered by the photodetector.

With the exception of Figure \ref{fig:comparison}, all data was taken by irradiating the cell walls more or less uniformly using a commercial blue LED (Optotech OT16-5100-RB, $\lambda_{LIAD}=455 \pm 20 \unit{nm}$, and $480\unit{mW}$ measured maximum output power).

\section{Data Analysis} \label{sec:theory}

A typical set of loading curves is shown in Figure \ref{fig:fitting} and demonstrates the photon stimulated desorption of sodium from the cell walls.  At $t = $ 0 the MOT current is turned on.  Subsequently, the number of trapped atoms increases and reaches a steady state whose value depends monotonically upon the LIAD intensity.  These curves can be analyzed in terms of a simple rate equation model for the number of trapped sodium atoms $N(t)$ \cite{monr90}
\begin{align}
\dot{N}(t)=R-(\gamma_\mathrm{Na} + \gamma_\mathrm{bg})N(t) \label{eq:original}
\end{align}
where $R$ is the MOT loading rate, $\gamma_\mathrm{Na}$ and $\colbg$ are the trap loss rates due to two-body collisions between the captured atoms with fast untrapped Na atoms and with background gases, respectively.  We have neglected intra-trap light-assisted collisions which appear as an additional quadratic term in the sodium density and did not play any significant role under our experimental conditions.  The exponential solution to Eq.\ \ref{eq:original} for $t>0$ is $N(t) = N_{\infty}(1-e^{-\coltot t})$, where $\coltot \equiv \colNa+\colbg$ is the total trap loss rate.  Each loading curve was fit to this functional form to extract $N_{\infty}$ and $\coltot$.  The initial loading rate is then determined as the product of the two extracted parameters $R = N_{\infty} \coltot$.

\begin{figure}[]
\centering
\includegraphics[width=0.5\textwidth]{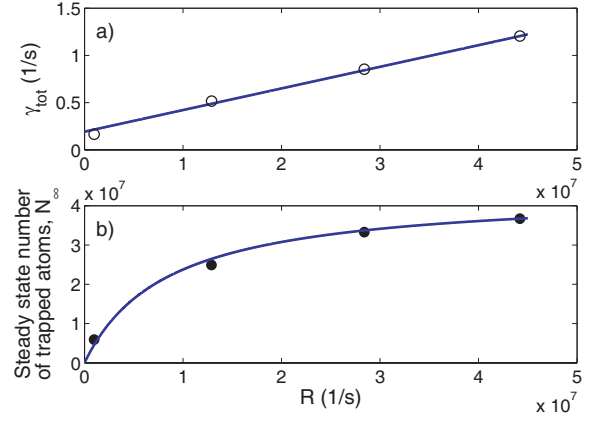}
\caption{Atom number and loading rate dependencies. By fitting each curve in Fig. \ref{fig:fitting} to the solution of Eq.\ (\ref{eq:original}), the total loss rate $\coltot$ (a), and the total number of captured atoms $N_{\infty}$ (b) are determined and plotted against the loading rate $R = \coltot N_{\infty}$. The solid lines are fits from the Eqns.\ (\ref{eq:params}) and (\ref{eq:params2}).}
\label{fig:fitting2}
\end{figure}

As elucidated by \cite{monr90}, for a vapor cell MOT both the capture rate $R$ and the loss rate $\colNa$ are proportional to the density of sodium and therefore to each other.  Thus we can parameterize $R = \alpha \colNa$ for some $\alpha$.  From this the extracted fitting parameters take on specific functional forms
\begin{equation}
\coltot = \frac{R}{\alpha} + \colbg
\label{eq:params}
\end{equation}
and
\begin{equation}
N_{\infty} = \frac{R}{\coltot} = \frac{\alpha R}{R+\alpha \colbg}
\label{eq:params2}
\end{equation}
In Figure \ref{fig:fitting2}, we test the validity of the above equations by plotting the experimentally determined value of $\coltot$ and $N_{\infty}$ versus $R$, which was varied by changing the intensity of the LIAD using the blue LED. In Fig.\ref{fig:fitting2}(a), the data clearly shows a linear relationship between $\coltot$ and $R$ whose slope yields $\alpha = 4.4\times10^7$, and whose intercept yields $\colbg = 0.19 \unit{s^{-1}}$.  Moreover, the steady state atom number fits well to Eqn.\ (\ref{eq:params2}) with the same value of $\alpha$ and $\colbg$ within statistical uncertainty.

The parameter $\alpha$ represents an upper bound for the atom number in the MOT.  In the limit where sodium collisions dominate the MOT loss rates,
$N_{\infty} \rightarrow R/\colNa = \alpha$ becomes independent of vapor pressure and depends only on the ratio of thermal to capture velocities for the atom, the volume enclosed by the trapping beams and upon other optical parameters (laser intensity, detunings and repumping sideband strength). In this situation, further increasing $R$ by turning up the light intensity used for LIAD will only shorten the time taken to reach steady state, not increasing the maximum trapped atom number. In Figure \ref{fig:fitting} one can clearly see that for high LIAD intensity $N_{\infty}$, which appears to be saturating near $4 \times 10^7$.  The inset shows, however, that the loading rate $R$ continues to increase linearly with LIAD intensity up to at least $4\unit{mW/cm^2}$.  The maximum measured $R$ was almost 50 times larger than in the absence of LIAD, while the steady state atom number was a little less than 10 times larger.

\subsection{The Role of Background Gas}
\begin{figure}[htp]
\centering
\includegraphics[width=0.5\textwidth]{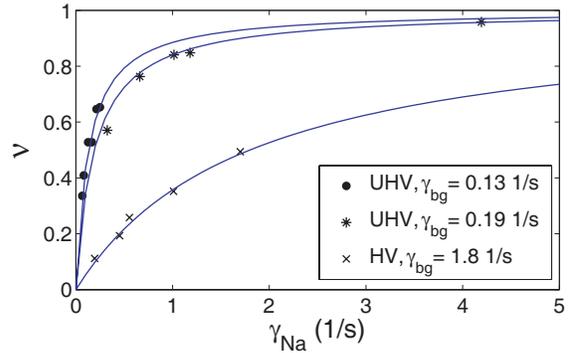}
\caption{Influence of background gas collisions. We present three data sets of the normalized atom number $\nu$ for two different vacuum regimes. The two UHV data sets were obtained by varying the blue LED voltage after different elapsed times since the Na dispenser was fired, while the HV data set was obtained by changing the dispenser current. The solid curves result from the direct use of Eqn.\ (\ref{eq:saturation}), with $\colbg$ determined by Eqn.\ (\ref{eq:params}).}
\label{fig:saturation}
\end{figure}

\begin{figure}[!t]
\centering
\includegraphics[width=0.5\textwidth]{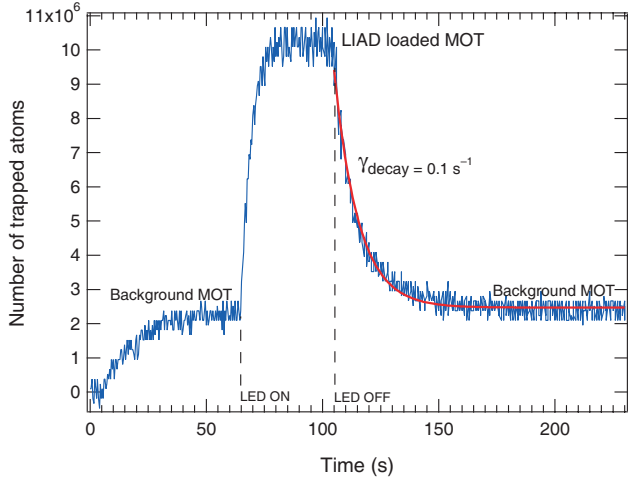}
\caption{LIAD loaded MOT is compatible with long lifetimes. The entire loading and decaying dynamics of the MOT with the blue LED LIAD is shown. The LED was turned on at time $t=65\unit{s}$, and turned off at $t=105\unit{s}$. By fitting the remaining trapped atom number for $t>105\unit{s}$ we extracted the loss rate of $0.10 \unit{s^{-1}}$, which is comparable to the value inferred from the steady state properties of the model Eqn.\ (\ref{eq:original}), as well as with the background MOT loaded previously, between 5 and $65\unit{s}$.}
\label{fig:decay}
\end{figure}

To further confirm the model of \cite{monr90} and clarify the role of background gas collisions, it is useful to normalize the steady state atom number to its maximum value $\alpha$.  If we define a dimensionless variable $\nu = N_\infty/\alpha$ it can be expressed via Eqn.\ (\ref{eq:params2}) as
\begin{equation}
\nu = \frac{\colNa }{ \colNa + \colbg} \label{eq:saturation}
\end{equation}
$\colNa$ is proportional to the density of sodium which can be varied by changing the LIAD intensity or the dispenser current. Therefore, $\nu$ runs from 0, when no sodium is present (in the background vapor), to 1, when collisions with Na completely dominate the MOT losses. The crossover between these two regimes occurs when $\colNa = \colbg$, the background gas collision rate. We changed the vacuum regime in order to study the dependence of this parameter on the MOT loading dynamics.

For both high vacuum (HV) and ultra-high vacuum (UHV) conditions, loading curves were acquired and analyzed in a manner similar to Figure \ref{fig:fitting}. Thus $\alpha$ was computed and both $\nu$ and $\colNa = R/\alpha$ could be determined experimentally.
The result is plotted in Figure \ref{fig:saturation}. We observed a dramatic dependence of the measured quantities on the vacuum conditions. The data sets all obey the same functional form given by Eqn.\ (\ref{eq:saturation}), but the crossover point occurs for much lower $\colNa$, i.e., sodium vapor density, for UHV when compared to HV.

From a fit to Eqn.\ (\ref{eq:params}) we can extract the values of the background gas collision rate. Under HV conditions where the pressure was $6 \times 10^{-9}\unit{torr}$, we obtained $\colbg \approx 1.8 \unit{s^{-1}}$. Under UHV conditions (pressure below $1 \times 10^{-11}\unit{torr}$), the background gas collision rate was found to vary depending on the time elapsed since the last firing of the dispenser.  The UHV curve with $\colbg = 0.19 \unit{s^{-1}}$ was taken a few hours after the firing whereas the curve with $\colbg = 0.13 \unit{s^{-1}}$ was taken a few months after the last firing. The ratio ($\colbg$)$_{HV}$/($\colbg$)$_{UHV}$ $\approx$ 10. By contrast, the vacuum gauge measured a pressure ratio of about 600. This is not an uncommon feature of UHV systems: due to mounting constraints the vacuum gauge had to be installed closer to the pumps where the pressure is typically lower than at any other location in the vacuum chamber, including the atom trap. In fact, the data in Figure \ref{fig:fitting2}(a) is more reliable to measure the lifetime of the trapped atoms than the pressure gauge readings.

The MOT saturates at fairly low sodium vapor density in the UHV regime, and it is typically larger and brighter. Under UHV conditions beyond the crossover point, increasing the vapor density by a factor of 5 increased the MOT atom number by only 14\%. This observation demonstrates that very low Na background vapor density is needed to completely fill the MOT at an appreciable atom number of $3.7 \times 10^7$.  Moreover, the UHV MOT did not require the dispenser to be fired over the course of many months, when running it no more than three to four times a week \cite{aubin05}.

The collisional loss rates in the range $\coltot = 0.1$ to $0.2\unit{s^{-1}}$, inferred from the data in Fig.\ref{fig:saturation}, suggest that the LIAD loaded MOT should be compatible with evaporative cooling to BEC.  We have also measured the time constant directly by observing the MOT dynamics.  The result is plotted in Fig.\ref{fig:decay}, where we recorded the time evolution of the atom number before, during and after the LED was switched off.  In that data set, the MOT atom number decayed from a peak of $10^7$ around $t = 105 \unit{s}$ to an asymptotic value of about $2.5 \times 10^6$ atoms with a $1/e$ time constant of $\tau=10\unit{s}$.

In the absence of LIAD, a small background Na partial pressure exists, allowing for a small MOT to be filled before $t = 65 \unit{s}$.  When the LED is turned on it creates a transient Na vapor pressure of much higher density that is responsible for the spike in the trapped atom number.  Upon shutting off the LED the extra vapor becomes quickly extinguished by being adsorbed onto the cell walls.  Collisions with this background sodium vapor as well as residual (non-sodium) vapor are the cause of the observed 10 second decay time.

\subsection{Mechanism of desorption}

According to Yakshinskiiy and Madey \cite{yakshinskiy99}, the photo-desorption of neutral sodium is a charge transfer process for Na$^+$ ions bound to the silica (SiO$_2$) surface. An electron can be excited from SiO$_2$ by photon bombardment. Above a threshold energy this electron transfers to the unfilled Na 3s level. The neutral Na atom then finds itself on a repulsive potential causing it to quickly dissociate from the surface. Previous work demonstrated a steep dependence of the neutral sodium yield on photon energy in the range of a few eV using a mercury arc lamp. However, the large error bars caused by the
broad wavelength spectrum of the mercury lamp precluded a more precise determination of the wavelength dependence in the region of the threshold. Fig.1 of reference \cite{yakshinskiy99} shows an error bar spanning the range of approximately $2.0-3.8\unit{eV}$ for the lowest energy filter setting. No additional information is available within or below that energy range as the experiment could not detect any appreciable signal. For a more precise determination of the desorption due to solar radiation incident on planetary bodies such as the Moon, it is the photon flux from the visible and ultraviolet radiation that plays the most important role. Therefore there is a clear need to examine this region in greater detail.

To address this issue we recorded the MOT loading curves using eight independent LEDs whose wavelengths ranged from $590\unit{nm}$ to $365\unit{nm}$. A black paper mask was custom built to fit over a small part of the Pyrex vacuum cell with a $2$ by $2\unit{cm}$ aperture on both front and back surfaces. This approach allowed us to precisely define and control the area exposed to the LIAD photons. Before the runs, the LEDs were individually adjusted to provide $2\unit{mW/cm}^{2}$ of light irradiance through the mask aperture. This was experimentally found to optimize the LIAD signal to noise ratio over the range of wavelengths used.  Since the spatial properties of each LED were slightly different, we carefully measured their individual intensity profile within the exposed area and applied an area average correction factor to the data.  The photon energies and energy uncertainties are provided by the manufacturer's specifications sheets \cite{leds1,leds2}.

To acquire the data, each run consisted of two full MOT loading curves, similar to those in Fig.\ref{fig:fitting}, taken in two steps. In the first step, the LED was kept off and the MOT was fully loaded from the background sodium vapor for a period of about $40\unit{s}$.  The MOT current was then immediately turned off and the LED was turned on and allowed to equilibrate thermally for about $20\unit{s}$.  In the next step the MOT was reloaded with the LED on for $40\unit{s}$.  Thus, in the first step, the MOT was loaded from the background Na vapor alone, while in the second step both background vapor and photo-desorbed atoms were present.  The final result was then computed as the difference between the rates $dN/dt|_{t=0}$ measured with LEDs on and off, yielding the net loading rate due to photo-desorption.

\begin{figure}[!htp]
\centering
\includegraphics[width=0.5\textwidth]{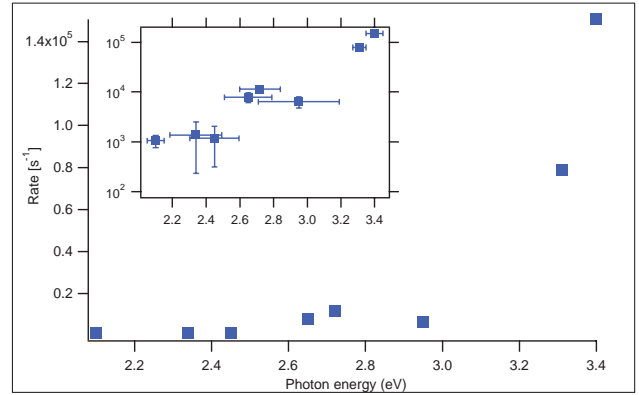}
\caption{Photon energy dependence of desorption of sodium atoms from Pyrex glass. Plotted is the excess MOT loading rate due to LIAD for an average intensity of $2\unit{mW/cm}^{2}$.  Each point is the average of 3-4 MOT loading curves. The inset shows the same data plotted on a logarithmic scale with the $y$-error bars resulting from statistical data analysis. The energy error bars correspond to the typical half-width uncertainties directly provided by the manufacturer's specifications sheets.}
\label{fig:comparison}
\end{figure}

The results are shown in Figure \ref{fig:comparison}, where one can clearly see that the desorption rates become significant only for photon energies above $2.6\unit{eV}$ ($470\unit{nm}$). For longer wavelengths we are unable to detect any significant LIAD effect, consistent with a threshold phenomenon. In the absence of a theoretical model we have refrained from fitting our data to any functional form. Nevertheless, we are able to detect a significant sodium desorption in the range $2.6$ to $3.4\unit{eV}$, a region below the detection threshold of previous work \cite{yakshinskiy99}. Part of the improvement lies in the use of LEDs rather than a Hg lamp, which has narrowed the spectral bandwidth associated with each data point by an order of magnitude on average. More significantly, however, our atom trapping technique is very sensitive to small numbers of desorbed atoms. For example, the majority of the data we presented in Figures \ref{fig:fitting}-\ref{fig:decay} were taken using light at $455\unit{nm}$ ($2.72\unit{eV}$), very close to the threshold observed for the LIAD effect on sodium. Therefore, our data on Pyrex complement the studies on pure SiO$_2$ surfaces \cite{yakshinskiy99,yakshinskiy03}, and there could be variations in the material surface properties that may affect the PSD as well.

The MOT loading rate depends upon both the photo-desorption cross-section, as well as the MOT capture efficiency. If the desorbed atoms re-thermalize with the cell walls before they enter the MOT laser beams, as we believe to be the case for our experimental scenario, the latter should be independent of the wavelength of the desorbing light. In this limit the loading rate of Figure \ref{fig:comparison} becomes a relative measure of the photo-desorption cross-section. However, it is important to note that the photo-desorption process does not necessarily need to produce a Maxwell-Boltzmann distribution of velocities, and the MOT capture efficiency need not be wavelength independent, as is often assumed to be the case.

Larger photon energies would increase the desorption rate dramatically, an advantage when producing large MOTs very quickly; however, shorter wavelengths will be significantly attenuated through the Pyrex walls. Future work could explore the difference between glass and quartz, although it is questionable whether the latter will be as useful for producing larger MOTs \cite{shengwang04}. We also plan to examine porous materials more closely resembling the lunar surface.

\section{Conclusion}

We have demonstrated that LIAD can be used effectively for fast loading of a sodium vapor-cell based magneto-optical trap, allowing for rapid control over the alkali background vapor pressure inside the cell. The same desorption process plays an important role in planetary science, albeit at drastically different temperatures from a laboratory BEC experiment.  Studies by the surface science community have increasingly begun to quantify radiation-induced chemical changes \cite{yakshinskiy99,yakshinskiy03,dominguez04}. The high sensitivity (potentially single atom) of magneto-optical traps may be a powerful tool for surface science.

C.R. thank Thomas Orlando for useful discussions. This work was supported by the U.S. Department of Energy, the Army Research Office and the National Science Foundation.

\bibliographystyle{apsrev}
\bibliography{References}
\end{document}